# Exploring the roles of ICT in supporting sustainability practices

**Abdon Carrera Rivera**
Melbourne School of Information Systems
The University of Melbourne, Australia
Email: acarrera@unimelb.edu.au

**Sherah Kurnia**
Department of Computing and Information Systems
The University of Melbourne, Australia
Email: sherahk@unimelb.edu.au

## Abstract

The concern about sustainability has arisen due to the overuse of natural resources and the increased use of energy consumption over the last decades. Information communication technologies (ICT) has the potential to address the three main aspects of sustainability (people, planet, profit) and therefore, several organizations have initiated a sustainable development by integrating ICT within their business activities. However, the roles of ICT in supporting sustainability initiatives have only been discussed in a limited number of studies and there is a lack of practical examples that demonstrate how the different roles of ICT are played out in an organization's environment. Therefore, this research aims to explore how ICT can be used by organizations to support sustainability initiatives. In particular, in this research-in-progress paper, we examine how a leading organization deploys Internet-of-Things as an example of an ICT application to support various sustainability initiatives. The study findings enhance the current understanding of how ICT can support sustainability practices of organizations.

## 1  Introduction

The rapid growth of business enterprises has contributed to the degradation of the natural environment due to the increased use of power consumption and overuse of scarce resources in the last decade (Elliot and Binney, 2008; Hedwig et al., 2009). The main sustainable development challenge comes from unsustainable consumption and production that have evolved over time. For example, human daily activities and needs such as demand for food, energy, increased human populations, production and consumption of materials and products in our daily lives have increased the amount of greenhouse gas concentrations by 20-40% in our atmosphere (Economic and Social Survey, 2013). This issue results in huge economic and social costs that could affect the welfare of the environment.

The environmental awareness has inspired companies to change their daily activities into more environmental friendly practices focusing on the sustainable development. The concept of sustainable development is based on a long term optimization of economical, ecological and social objectives to generate a lasting performance on the organization activities. These three aspects (economical, ecological and social) are known as the Triple Bottom Line (TBL) (Elkington, 1997). Some organizations have planned sustainable strategies that not only minimize the environmental impact but also increase their business competitiveness (Porter and Kramer, 2011). One strategy is to find innovative ways to integrate/introduce information and communications technologies (ICTs) in the business processes in order to deliver sustainable benefits (Donnellan, Sheridan and Curry, 2011). ICTs include hardware, software, communication devices, network and so on, which enable users and organizations to capture, process, manipulate and share data and information. Over the years, ICT has provided organizations with capabilities to improve their profitability, efficiency, productivity and increase their competitive advantage. Thus, ICT has become a major supporter to business innovation, wealth generation and an important tool for organizations to develop sustainability capabilities (Elliot and Binney, 2008). From a sustainability view point, ICT enables firms to standardize, capture, analyse, monitor and utilize data (or information) that helps identify and change actions for the welfare of the community and planet through the development of new applications and systems. In addition, it helps reduce the energy use by dematerialization such as replacing physical data centres with Cloud computing (Melville, 2010; Erdmann et al., 2004). Enkvist, Naucler, and Rosander (2007) also estimate that ICT can provide business solutions to minimize its global greenhouse gas footprint impact.

Consequently, ICT can play significant roles in addressing sustainability (Melville, 2010; Rahim et al., 2014). Therefore, leading companies have implemented information communication technologies to fight against the negative environmental effect of the business processes and activities (Erek et al., 2009), improve their own economic benefit and offer benefits to the society (Hasan et al., 2009). However, limited research on the possible roles of ICT in supporting sustainability initiatives has been reported in the available sustainability literature (Dao et al., 2011; Elliot, 2011; Kurnia et al., 2012). A better understanding and practical examples of the roles of ICT from a sustainability perspective are still required. For this reason, in an effort to contribute to the IT/IS discipline and future studies to understand the potential of ICT in supporting sustainability, we aim to explore how ICT can support sustainability practices within an organization along the three dimensions of sustainability. The key research question addressed is:

*How are possible roles of ICT played out in supporting sustainability practices within organizations?*

In this research-in-progress paper, we investigate a published case of a leading company that has successfully deployed a specific ICT to support sustainability practices within their main business operations. Based on the four roles of ICT (Automation, Transformation, Informating and Infrastructure) within the sustainability framework proposed by Dao et al. (2011) and some key practices to address in the TBL identified by Kurnia et al. (2012), the paper explains how the roles of Internet-of-Things (IOT) can be played out in supporting sustainability initiatives of the case organization. Specifically, this research shows that the Informating and Infrastructure roles are played out in most business practices involved in the three pillars of sustainability. It further demonstrates that the Transformation role of ICT is critical in supporting initiative addressing the environmental aspect of sustainability.

## 2　Research method

This research-in-progress paper explores the experience of a successful organization in introducing an ICT to support the Triple Bottom Line based on published case study data. Case study is a suitable research method for this study which is exploratory in nature and addresses a 'how' question (Yin, 1984). Case study allows for a detailed investigation and understanding of the problem being investigated (Gable, 1994). After an extensive literature review, a number of publications related to how the selected case organisation, which is a leading organisation in introducing ICT to address sustainability were identified. Useful examples and insights into the changes in daily functions of the organisation in order to become a more sustainable organization by using different information communication technologies were available. Although there are other published cases regarding other organizations' sustainability practices available, none of them provide comprehensive details of practices and the use of ICT as the selected case. Furthermore, since there have been very limited existing studies with detailed descriptions of how ICTs are used to support diverse sustainability practices, the selected case organization is considered a revelatory case which enables us to conduct a preliminary investigation into a set of sustainability practices and how ICT supports them. The case company is referred in this paper as 'Genius Corporation'. Because of the comprehensiveness of the published data, it enabled us to analyse the case in detail to address the research question posed.

Content analysis technique is deployed to analyse the available data. Berg and Lune (2012, p. 349) define content analysis as "a careful, detailed, systematic examination and interpretation of a particular body of material in an effort to identify patterns, themes, biases, and meanings". We identified relevant organizational practices involved along the three dimensions of the TBL and identified patterns associated with the roles of ICT in supporting sustainability practices along the three aspects of TBL. In doing so, we apply the qualitative data analysis technique involving open, axial and selective coding (Mile and Huberman 1993). Although there are many ICT initiatives introduced by the case organisation, as a research in progress, this paper only focuses on Internet-of-Things (IoT) as an example of ICT to support sustainability initiative.

## 3　Key concepts and definitions

### 3.1　Sustainability

Sustainability can be considered as a means for survival since it encourages preservation of environmental, economic or social conditions for the sake of future generations and the use of resources to a degree where it is possible to restore them within a regeneration cycle (Erek et al., 2009). Therefore, many entities found that sustainability is an important business responsibility that affects the revenue generation of an organization, reputation, and cost reduction. The most accepted definition of sustainability comes from the Brundtland Commission which defines sustainability as a "development that meets the needs of the present without compromising the ability of future generations to meet their own needs" (WCED, 1987, p.43). This interpretation represents the core of sustainability and also has encouraged the elaboration of different concepts, such as corporate sustainability, based on its main idea. Although sustainability can be applied in many different disciplines, most studies agree that it is related to the conservation of environmental, social and economic dimensions for the welfare of future generations. The three dimensions represent the three main pillars of sustainability known as the Triple Bottom Line (TBL) (Elkington, 1997).

## 3.2 Triple Bottom Line (TBL)

The TBL is based on three main components: planet, people and profit (Elkington, 1997) with the idea to embrace the corporate sustainability objectives outlined in the Brundtland Report (WCED, 1987). The TBL principle highlights an important fact that organizations generate value in multiple dimensions, which are environmental, economic and social dimensions. Savitz (2006) explains that the TBL "captures the essence of sustainability by measuring the impact of an organization's activities in the world ... including both its profitability and shareholder values and its social, human and environmental capital."(p.15). Therefore, the main goal of a corporation in terms of sustainability is to address and enhance the triple bottom line performance. Elkington (2005) argues that the most appropriate way to ensure that an organization completely addresses the TBL is to incorporate all the principles of the three dimensions into their main business operations.

The economic dimension is the most popular aspect that has been traditionally addressed in most organizations. The economic aspect provides ways to add value to investments (investment opportunities), obtain good margin profits, and improve business performance and revenue (Ratan et al. 2010). The environmental or ecological dimension is elaborated by the consumptions of resources, which can be renewable and non-renewable, and the impacts on the ecosystem. Main activities that follow the concept of the environmental dimension in the TBL include efficient consumption of energy, proper disposal of toxic waste, appropriate use of renewable energy sources and natural resources. The social dimension can be considered as a supportive dimension which focuses on people and stakeholders inside and outside the organization and the management of social resources (Dreyer et al., 2006). Organizations follow the social dimension of the TBL to improve the welfare of their employees and the broader community.

## 3.3 Roles of ICT in sustainability

Currently, there exist limited studies that adequately explore how ICT can be used to enable new strategic business objectives which help organizations develop sustainability innovations and demonstrate a significant enhancement in their sustainability performance and competitiveness (Chen et al., 2008; Melville, 2010; Bengtsson and Agerfalk, 2011). To address the limitations with the existing ICT enabled sustainability studies, this work offers a better understanding of the different roles that ICT can play in supporting sustainability initiatives by exploring the insights and good practices of a leading company on how it deploys ICT to improve various practices aligned with the three pillars of sustainability.

Information and communications technology has been highlighted in many studies (Elliot and Binney, 2008; Melville, 2010; Hedwig et al., 2009) to be an important component in business daily activities and has shown a positive or negative effect to the environment. Due to its importance, organizations are recognizing their responsibility in environmental and social dimensions by exploring ICT- enabled sustainability practices that can lead to improvement in organizational sustainability performance.

This study adopts the four roles of IT/IS proposed by Dao et al. (2011) and a number of the key sustainable SCM practices (which can be extended to generic organization's practices) along the three dimensions of the TBL identified by Kurnia et al. (2012). As summarised in Table 1 Dao et al. (2011) defines 4 key roles of ICT resources which are: Automation, Informating, Transformation and Infrastructure. Each role plays an important function in different organizational practices and enables firms to develop sustainability capabilities. In terms of the Automation role, ICT have been used in many organizations to replace manual business processes with computerize application systems. With the Informating role, ICT is expected to help organizations produce useful information from the various data captured by the different IT/IS. Through the Transformation role, ICT can be used to re-engineer business operations and introduce significant changes into the way organizations carry on their business process. The Infrastructure role embraces hardware, software and network components to support business processes.

Kurnia et al. (2012) synthesize several SCM practices relevant for the dimensions of the TBL. Most of the practices described in Kurnia et al. (2012) form part of general organizational practices. For example, practices in terms of economic aspects include creating competitive advantages and enhancing sales. In terms of environmental dimensions, practices include clean/lean production (Manufacturing) and green distribution/logistics (Transportation), while practices from social dimension include Community relations and Work safety. All the practices mentioned in the example can be developed by any organization. As several studies have argued (Donnellan et al. 2011, Melville 2010), without information technologies, it is impractical to manage organization practices efficiently and effectively. However, the practices mentioned by Kurnia et al. (2012) are specifically relevant in the context of supply chain management. Therefore, in this study some practices identified have been revised to cover general business practices as shown in Table 1. Most practices listed are general enough for organizations to consider in addressing the three dimensions of sustainability. Table 1 also shows the possible roles that ICT can play in each practice, as conceptualized in Kurnia et al. (2012).

| Dimension | Key Practice | ICT Roles | | | |
|---|---|---|---|---|---|
| | | Automation | Informating | Transformation | Infrastructure |
| Economic | Reducing costs | Y | Y | | Y |
| | Achieving stakeholders satisfaction | | Y | | Y |
| | Enhancing sales | Y | Y | | Y |
| | Creating new business models | Y | Y | Y | Y |
| | Quality initiatives | | Y | | Y |
| | Creating competitive advantage | | Y | Y | Y |
| Environmental | Clean/Lean production (Manufacturing) | Y | Y | Y | Y |
| | Green Distribution/ Logistics (Transportation) | | Y | Y | Y |
| | Efficient resource consumption | | Y | Y | Y |
| Social | Community relations and communication | | Y | Y | Y |
| | Collaboration within company | | Y | Y | Y |
| | Employee satisfaction and well-being | | Y | Y | Y |
| | Education support | | Y | | Y |

*Table 1. Roles of ICT in supporting various sustainability practices (adapted from Kurnia et al. 2012)*

In order to understand how the roles of ICT are played out to support sustainability, this research in progress investigates in detail a specific information communication technology developed by a leading organization and explores the benefits of using ICT in different organizational operations to address key practices involved in the three pillars of sustainability.

## 4  Overview of the Case Study

Genius Corporation is one of the largest manufacturers of semiconductor chips with over 100,000 employees and annual revenue of approximately US $55.870 billions in 2014. In the last 10 years,

Genius has become an active company that promotes innovations and focuses on the creation and development of smart technologies with the goal of connecting devices to every person on earth. Genius is selected as a case study due to its recognition as a leading performer in the environmental sustainability of ICT and its long history of commitment to the environment.

The ICT application of Genius considered in this study is the Internet of Things. IoT is an example of an ICT application because it integrates billions of devices connected to the internet which enable organizations to rapidly store, access, manipulate and transmit data within the network. Most of those devices form part of everyday activities while others are integrated to daily functions. The technological advancement has contributed to the enhancement of the IoT capabilities by reducing the physical size of the devices but at the same time including more computational power, integrating connectivity with other devices and offering low power consumption. Therefore, most IT/IS practitioners find a huge potential in incorporating IoT with conventional information systems, business intelligence (BI), business analytics and business processes (Lee and Lee 2015). The power of IoT relies on the acquisition of data, analysing data and acting on that data. Having several devices being connected and communicating with each other offers significant advantage for business and societies. Because of the power of collecting data and analysing data, many decision systems can make better predictions and support better decision making activities based on the meaningful information obtained from IoT.

## 5  The Study Findings

In this section we show how the ICT (Internet-of-Things) implemented by Genius play different roles to support various sustainability initiatives. Furthermore, we highlight the key practices of the case organization that can be affected by the deployment of IoT. To evaluate the roles of the IoT, three main operation areas are considered. The operation areas include logistics/fleet management, manufacture/fabric process and data centres. Exploring these three main operation areas of a leading organization helps understand how ICT can enable sustainability practices within an organization and how the roles of ICT are played out to support the three dimensions of sustainability.

*Logistics and Fleet Management*
Table 2 summarizes key practices identified in the area of logistics and fleet management that are enhanced by the deployment of the IoT. In this particular operation area, all business practices show the use of IoT infrastructure to support sustainability initiatives. Hence, the Infrastructure role played out by the Internet-of-Things is embedded across all the various sustainability practices. The infrastructure includes smart wireless sensors, smart fuel meters, vehicle's console, gateways with analytics systems and cloud computing for communication between devices or systems.

| Dimension | Key practice | IoT role | Description |
|---|---|---|---|
| **Economic** | Reducing costs | Informating and Infrastructure | Fuel expenses reduction from the efficiently use of resources. |
| **Environmental** | Green Distribution/Logistics (Transportation) | Informating, Transformation and Infrastructure | Reduction of the air pollution and $CO_2$ emissions from vehicles. |
| **Social** | Education support | Informating and Infrastructure | Real time coaching about how to drive efficiently and information about the status of the vehicle. |
| | Work safety | Informating and Infrastructure | Prevention measures based on the diagnosis of the vehicle. |

*Table 2. Key practices enhanced by the IoT in the area of logistics and fleet management*

On the economic dimension, IoT helps Genius reduce costs through the Informating role. IoT can provide information about the potential savings in terms of fuel basis based on the efficiently use of fuel resources from the vehicles. Moreover, the data collected from wireless sensors can give an accurate estimation of the amount of fuel required for a distribution, resulting in reduction in fuel expenses.

On the environmental dimension, IoT demonstrates the Transformation role by helping Genius to perform green logistics distribution/transportation as a result of changing Genius' traditional logistics operation into a more resource efficient and environmental-friendly logistics distribution. This transformation of operations is achieved by using IoT to introduce a new efficient way of using fuel resources for distribution. The major change in logistics distribution/transportation is the capability to provide real time services such as tracking vehicles, monitoring fuel consumption and monitoring carbon emission of a vehicle. Genius has deployed different computing devices such as small wireless sensors and Bluetooth devices on different distribution vehicles. Each device monitors and collects data from the vehicle to a data centre. The raw data collected from the vehicle by the IoT can provide information about the fuel consumption of each vehicle. The Informating role is also played out by allowing drivers or logistics manager to receive the information gathered by the IoT from different vehicles. The information can be further analysed to achieve efficient fuel consumption, resulting in a reduction of the amount of $CO_2$ combustion emissions produced by the vehicles and reduction of air pollution.

Meanwhile on the social dimension, IoT improves the education support and work safety practices. The IoT Informating role is played out to support education by sending information to an edge device inside the vehicle (vehicle´s console) about how to drive safely and how to drive effectively so that drivers can minimize the fuel cost and air pollution. Also, the historical information gathered over time from IoT devices can give details about the driver's driving behaviour. With some analysis of the data, IoT can provide real time coaching on the shifting, braking and accelerating of the vehicle. In addition, through the Informating role of the IoT, work safety practice is improved by communicating and evaluating data from the IoT's sensors about the status of the vehicle (engine and tyres status). The data collected can give insights into the condition of the tyres of the vehicle and provide information regarding whether or not a tyre is correctly inflated or is correctly aligned with other tyres on the vehicle. Other analysis from the data may deduce how thin the rubber on the tyre might be and will give a prediction of how soon it will need to get replaced. Therefore, the benefits of deploying IoT include detailed diagnostic of the vehicle that gives information when something is wrong on the vehicle and communicate any potential safety incidents that can affect the driver's health & wellbeing.

*Manufacture and Fabric Processes*

| Dimension | Key practice | IoT role | Description |
|---|---|---|---|
| **Economic** | Reducing costs | Informating, Automation and Infrastructure | Cost expenses reduction from the efficiently use of resources and lower maintenance cost. |
| | Quality initiatives | Informating and Infrastructure | Improvement on the quality of production. |
| | Creating competitive advantage | Informating and Infrastructure | Manufacturing operations adapt faster to customer demands. |
| **Environmental** | Clean/lean manufacturing | Informating, Automation, Transformation and Infrastructure | Efficient use of resources by monitoring the energy usage on the factory and minimizing the energy consumption from machinery. |
| **Social** | Collaboration within company | Informating and Infrastructure | Engagement of employees with the manufacturing network using IoT |
| | Work safety | Informating and Infrastructure | Preventing measures about the potentially dangerous malfunction of equipment or machines |

*Table 3. Key practices enhanced by the IoT in the area of manufacture*

Table 3 summarizes the four different roles that IoT can play in supporting manufacturing sustainability practices. It shows that the Infrastructure role is played out in all the sustainability initiatives involved in the manufacturing operation area since most of the key practices supported by the IoT needs a physical or logical resources to deliver business objectives. The infrastructure includes big data analytics, decision systems, gateways with analytics, cloud computing, and wireless sensors with interconnected capabilities for each machine at the manufacturing process.

On the economic dimension, there are three key practices involved which include cost reduction, quality initiatives and creating competitive advantage. The cost reduction practice is supported through the Informating role of IoT. The historical data collected from wireless sensors and analytics can predict and communicate when the performance of manufacturing operations is decreasing. This offers benefits such as improved equipment performance, lower maintenance cost, reduced machine failure/downtime and cost savings of millions of dollars from Genius' manufacturing operation. Moreover, based on the sensor's information analysis, manufactures can take actions to increase the productivity of the factory, increase manufacturing margins and avoid unwanted costs.

The automation role is also played out to support cost reduction. IoT through sensors integrated in every physical manufacturing tool enables the communication between machinery and business systems. The communication of manufacturing machines and business systems (machine-machine communication) allows manufactures to optimize their production without human intervention. This results in less manual interactions and errors, leading to a reduction in manufacturing cost and increased productivity.

IoT helps Genius to support quality initiatives through the Informating role by extracting valuable insights from industrial data such as how well the manufacturing process is running. For example, on an assembly manufacturing factory, there are millions of units that are produced for customers and there are thousands of machines that assemble and produce the final product. The IoT data analytics ensure that factories are producing products with the highest quality and aligned with industry standards by keeping a clear visibility of the factory flow and processes. IoT also improves Genius' ability of creating competitive advantage by playing the Informating role. The amount of data that flows in a factory through the IoT creates new opportunities for predicting demand and making better decisions to response to changing market's demands, which creates new economic value. Due to the complexity of manufacturing operations Genius used IoT to communicate and improve the efficiency in their assembly factories making faster and precise decisions with real-time indicators and enabling operations to be adapted based on the customer's demand.

On the environmental dimension, four roles of IoT are observed in supporting clean/lean manufacturing (production) which include Infrastructure, Automation, Informating and Transformation. The main role that IoT can play to support clean manufacturing is the Transformation role. Genius has transformed its factories from traditional manufacturing operations into clean, environmental friendly and smart manufacturing operations (Intel, 2014b). With the introduction and deployment of IoT in factories, Genius has transformed its manufacturing process to include several monitoring capabilities through environmental sensors that enables efficient use of resources and reduction in waste generation. Genius has changed its manufacturing process by integrating smart wireless sensors into different manufacturing machines to adjust their operation based on the data analysed, and to reduce waste, energy consumption as well as the emission of pollutants.

The Automation role is also played out by integrating intelligence across the factory. Smart sensors embedded in the manufacture tools allow machines to communicate, make decisions and take actions automatically based on what they have sensed. Through the Informating role, the IoT's sensors enable monitoring and informing the energy use from all tools and machines involved in the manufacturing process. The data analysed from the sensors can provide a more efficient way to manage resources in a factory, resulting in reduced energy consumption that can minimize negative effects on the environment. For example, during Genius' assembly process, IoT sensors can detect in real-time if any equipment starts to demand more energy than usual. The information helps to alert and take actions to improve the use of resources.

On the social dimension, the Informating role of IoT is played out to support collaboration and communication within the company. IoT keeps employees more informed by connecting all manufacturing tools and creating a network that improves employees' participation and engagement with the development of new manufacturing capabilities using the IoT. The IoT demonstrates the Informating role to perform work safety practices. The implementation of IoT and the intelligence provided by the big data analytics, allows the company to obtain better knowledge from several computing devices about any manufacturing problems that cause machines to fail, behave incorrectly and detect processes that are getting out of control. The power of sensing data from manufacturing tools also identifies if any equipment/machine is old, become useless or is potentially dangerous in the manufacturing process. Therefore, Genius has allocated several wireless sensors in different machines so that they can give better insights into the root cause of manufacturing issues and malfunctioning tools that can improve employee's well-being.

*Data Centres*
The last operation investigate is data centres of Genius. Table 4 shows four relevant key practices identified along the TBL dimensions that are influenced by the use of IoT.

| Dimension | Key practice | IoT role | Description |
| --- | --- | --- | --- |
| **Economic** | Reducing costs | Informating and Infrastructure | Cost expenses reduction from the efficiently use of resources and adequate server management. |
| | Creating competitive advantage | Informating and Infrastructure | Faster adaption of physical infrastructure to meet new demands on the systems. |
| **Environmental** | Efficient resource consumption | Informating, Transformation and Infrastructure | IoT provides efficient ways to minimize the resource consumption resulting in a reduction of greenhouse gases. |
| **Social** | Work safety | Informating and Infrastructure | IoT creates awareness and alert managers potential issues that can arise in the data centre. |

*Table 4. Key practices enhanced by the IoT in the area of data centre*

One of the issues of data centres from an environmental point of view is that their infrastructure and operations directly impact the environment by generating tons of $CO_2$. Therefore, in an attempt to reduce the environmental damage created by the data centres and at the same time manage them efficiently without reducing the performance, Genius implemented and deployed IoT on the data centre environment. Table 4 also shows that the Infrastructure role played out by the IoT supports all the sustainability initiatives on the data centre operations. The Infrastructure role in this particular operation area is based on tools such as environmental sensors (air pressure, temperature, energy and humidity sensors), smart meters, data analytics and cloud computing.

On the economic dimension, IoT shows the Informating role through the ability to help Genius to reduce costs. The reduction of resources consumption directly affects the operational costs. Hence, an efficient management of resources brings a significant reduction in business energy costs. Genius uses IoT smart energy metering to help reduce operational costs by monitoring and adjusting possible efficient ways to use environmental resources. The competitive advantage practice is also supported by the IoT playing the Informating role. The power of the data analytics embedded with the IoT also helps to inform and predict future growing infrastructure on the data centre according to the current demand of systems, resulting in a powerful advantage from other business. Moreover, the analysis from the sensor's data helps managers to take better decisions and to dynamically adapt the data centre infrastructure to meet new business demands.

On the environmental dimension, IoT helps Genius to support efficient resource consumption by playing the Transformation role. IoT has transformed Genius' data centre operation by bringing a different way to manage the data centre environment. IoT provides a new way to obtain a clear visibility of the data centre infrastructure which monitors the energy and electricity requirements to power the data centre workplace, the water required for cooling server racks and data centre environment conditions such as air-flow, temperature, air pressure and humidity. Therefore, Genius changed part of its data centre infrastructure to include and deploy IoT's sensors combined with data analysis for an efficiently management of resources in the data centre environment. The Informating role is also played out to support efficient resource consumption by using wireless sensors as a monitoring and information platform that collects data from data centre environment and gives efficient analysis of the correct use of resources such as water, energy and power from servers, resulting in efficient way to minimize environmental impacts, reduce $CO_2$ and greenhouse gases emissions.

Meanwhile, on the social dimension, IoT improves work safety within company through the Informating role. The information collected from IoT can inform managers of possible problems such as presence of water leaks, open cabinet doors or smoke inside the data centre infrastructure. The environmental sensor gives alarms to managers about any status of the data centre so they can proactively manage the possible issues that can arise.

# 6   Discussion and Conclusion

Many business organizations have recognized the potential role that ICT can play in enabling and supporting sustainability. However, limited studies have discussed the roles of ICT in supporting sustainability initiatives and very few studies have shown practical examples of the role that ICT can play to support the three pillars of sustainability within a specific organization. This preliminary study provides useful practical examples to show how ICT can be used to support sustainability practices within organizations and clearly illustrate the 4 roles of ICT which are currently lacking in previous studies. Kurnia et al. (2012) only provides a conceptual analysis of the various roles that ICT can play in supporting sustainability practices. This study has extended Kurnia et al. (2012) study by investigating a specific application of ICT (Internet-of-Things) and illustrates the various roles it can play in supporting different business practices. Moreover, this study has shown that ICT can play a significant role in supporting sustainability practices contrasting previous studies which consider ICT as a contributor of sustainability issues (Hedwig et al., 2009; GAO, 2005; Elliot and Binney, 2008).

This research in progress has demonstrated possible ways ICT can be used to help an organization address the TBL. The evaluation of the possible roles of ICT indicates that Informating and Infrastructure roles are the most widely applicable role because in every IT or IS there must be data and information involved in business processes. Thus, an ICT is required to analyse, process, manage and communicate the information available to achieve business objectives. Moreover, there must be an infrastructure (physical or virtual) that supports the information to deliver business objectives. This study found that the Infrastructure role is played out through the provision of wireless sensors embedded across the different operation's environment. The Informating role is played out by using the data collected from the sensors to extract or analyse meaningful information. The Automation role is fulfilled through machine-to-machine communications through sensor wireless networks that reduce human intervention and automate business processes, especially supporting environmental and economic aspects. One interesting finding is the importance of the Transformation role in supporting environmental dimensions. This finding confirms that any organization that wants to address and support environmental aspect should re-engineer their business operations to adopt practices aligned with environmental dimensions. The IoT discussed in this study provides evidence to show how it can be used to transform some organizational aspects on three main operation areas at Genius, resulting in more intelligent, efficient and environmental-friendly operations.

This study has demonstrated that the most widely applicable roles of ICT to support sustainability initiatives are Informating and Infrastructure. Moreover, the research highlights the importance of the Transformation role along the environmental dimension and the support it offers to main business operations of an organization. This study has contributed to the IT/IS literature by providing practical evidence of how ICT can play different roles in supporting sustainability initiatives to enhance the current understanding in this area. Since practice leads theory in the area of sustainability and ICT, by providing detailed examples of how ICT can support sustainability practice within a manufacturing organization, the study findings enhance the current understanding and complement the limited literature in this area. The understanding obtained in this study will contribute to future theoretical development regarding effective application of ICT to support sustainability practices. Furthermore, this study offers practical contributions by giving insights into how a particular ICT such as Internet-of-Things can help organizations improve their sustainability business practices and showing the benefits of the development and deployment of Internet-of-Things in three main operation areas of a leading organization.

In this research-in-progress paper, our work is only based on published case study data. Although it is useful, future studies assessing the real context through in-depth case studies will be required to gain deeper understanding of how ICT support sustainability practices in various ways. Moreover, this study has only explored in detail a single information technology (Internet-of-Things) under three main operation areas (manufacture, logistics and data centre). Therefore, in our future research involving a multiple case study, we will explore several widely used ICTs to obtain a deeper insight into other possible roles that ICT can play in different contexts. Such in-depth case studies will further contribute to the current understanding of how organizations can deploy ICT to innovate and transform their business practices to enhance economic, environmental and social performance.